\documentstyle[amssymb,preprint,aps]{revtex}

\begin{document}
\title{Gravitomagnetic time delay and the Lense-Thirring effect in Brans-Dicke
theory of gravity }
\author{A. Barros$^{a}$ and C. Romero$^{b}$}
\address{$^{a}$Departamento de F\'{i}sica, Universidade Federal de Roraima,\\
69310-270, Boa Vista, RR, Brazil\\
$^{b}$Departamento de F\'{i}sica, Universidade Federal da Para\'{i}ba, Caixa%
\\
Postal 5008, 58051-970, Jo\~{a}o Pessoa, PB, Brazil\\
e-mail: cromero@fisica.ufpb.br}
\maketitle

\begin{abstract}
We discuss the gravitomagnetic time delay and the Lense-Thirring effect in
the context of Brans-Dicke theory of gravity. We compare the theoretical
results obtained with those predicted by general relativity.We show that
within the accurarcy of experiments designed to measure these effects both
theories predict essentially the same result.
\end{abstract}

\section{Introduction}

\bigskip The idea that mass currents generate a field called, by analogy
with electromagnetism, the gravitomagnetic field, is no doubt a very
appealing one\cite{Lense}. As is well known, according to general
relativity, moving or rotating matter should produce a contribution to the
gravitational field that is the analogue of the magnetic field of a moving
charge or magnetic dipole. A clear manifestation of this field may be found,
for instance, in the Lense-Thirring precession\cite{Wheeler}, an effect
predicted soon after Einstein formulated general relativity. The
gravitomagnetic field also contributes to the gravitational time delay\cite
{ciufolini}, among other effects.

The real possibility that gravitomagnetic effects can be measured with the
current technology of laser ranged satellites (LAGEOS and LAGEOS II) has
aroused great interest in the subject\cite{lageos}. It should be mentioned
that the Relativity Gyroscope Experiment\ (Gravity Probe B) at Stanford
University, in collaboration with NASA and Lockheed-Martin Corporation, has
a program of developing a space mission to detect gravitomagnetism effects
directly. Certainly, these experimental programs will open new possibilities
of testing general relativity against other metric theories of gravity \cite
{Camacho,Will}, in particular the scalar-tensor theory, one of the most
popular alternatives to Einstein theory of gravitation. Our aim in this
paper is to investigate two gravitomagnetic effects: the time delay and the
Lense-Thirring precession in the context of Brans-Dicke theory, and then
compare the results with those predicted by general relativity.

\ To define the gravitomagnetic field in general relativity one assumes the
weak field and slow motion approximation. On the other hand, it has been
shown that in this approximation solutions of Brans-Dicke equations are
simply related to the solutions of general relativity equations for the same
matter distribution\cite{Barros}. By using this fact one can easily
establish a straightforward correspondence between weak field effects in
both theories, in particular gravitomagnetic effects.

This paper is organized as follows. In Section II, we give a brief
introduction to the basic ideas of gravitomagnetism. Then, in Section III,
we show how general relativity and Brans-Dicke theory of gravity are related
in the weak field approximation. The gravitomagnetic field in Brans-Dicke
theory is defined in Section IV. We consider the Lense-Thirring effect and
the gravitomagnetic time delay in Brans-Dicke theory in Sections V and VI,
respectively. Section VII is devoted to some remarks.

\section{The gravitomagnetic field in general relativity \ \ }

Let us recall that in the weak field approximation of general relativity we
assume that the metric tensor $g_{\mu \nu }$ deviates only slightly from the
flat spacetime metric tensor. In other words, we assume that $g_{\mu \nu
}=\eta _{\mu \nu }+h_{\mu \upsilon }$, where $\eta _{\mu \nu }=diag(-1,1,1,1)
$ denotes Minkowski metric tensor and $h_{\mu \upsilon }$ is a small
perturbation term. Then, by keeping only first-order \ terms in $h_{\mu
\upsilon }$, the Einstein equations become

\begin{equation}
\square \overline{h}_{\mu \nu }=-\frac{16\pi G}{c^{4}}T_{\mu \nu }
\label{weakfieldgr}
\end{equation}
where $\overline{h}_{\nu }^{\mu }=h_{\upsilon }^{\mu }-\frac{1}{2}\delta
_{\nu }^{\mu }h$ and we are adopting the usual harmonic coordinate gauge $%
\left( h_{\upsilon }^{\mu }-\frac{1}{2}\delta _{\nu }^{\mu }h\right) ,_{\mu
}=0$.

If we assume a non-relativistic matter distribution with a mass density $%
\rho $ and velocity field $\overrightarrow{v}$ , then (\ref{weakfieldgr})
yields 
\begin{equation}
\square \overline{h}_{00}=-\frac{16\pi G}{c^{2}}\rho 
\label{gravitoelectric}
\end{equation}
\begin{equation}
\square \overline{h}_{0i}=\frac{16\pi G}{c^{3}}\rho v_{i}
\label{gravitomagnetic}
\end{equation}
where $v_{i\text{ }}$denotes the velocity components, and terms such as $p$
and $v_{i}v_{j}/c^{4}$ have been neglected. Let us now specialise  to the
case of a stationary gravitational field of a slowly rotating body. Then (%
\ref{gravitoelectric}) and \ (\ref{gravitomagnetic}) reduce to  
\begin{equation}
\nabla ^{2}\left( \frac{c^{2}\overline{h}_{00}}{4}\right) \equiv \nabla
^{2}(\Phi _{g})=-4\pi G\rho   \label{ro}
\end{equation}
\begin{equation}
\nabla ^{2}\overline{h}_{0i}=\frac{16\pi G}{c^{3}}\rho v_{i}  \label{ui}
\end{equation}
where $\Phi _{g}$ is the gravitoelectric scalar potential (the gravitational
counterpart of the electromagnetic scalar potential). Far from the source we
will have 
\begin{equation}
\Phi _{g}=\frac{GM}{r}  \label{fi}
\end{equation}
\begin{equation}
\overrightarrow{\overline{h}}=-\frac{2G(\overrightarrow{J}\times 
\overrightarrow{r})}{c^{3}r^{3}}\equiv -\frac{2\overrightarrow{A}_{g}}{c^{2}}
\label{ag}
\end{equation}
where $\overrightarrow{A}_{g}$ is the gravitomagnetic vector potential
vector ( the gravitational counterpart of the electromagnetic vector
potential), $\overline{h}_{0i}$ are the components of the vector $%
\overrightarrow{\overline{h}}$, $M$ and $\overrightarrow{J}$ are the total
mass and angular momentum of the source. In close analogy with
electrodynamics we define the gravitoelectric field to be $\overrightarrow{%
E_{g}}=-{\bf \nabla }\Phi _{g}$ and the gravitomagnetic field to be 
\begin{equation}
\overrightarrow{B}_{g}=\overrightarrow{{\bf \nabla }}\times \overrightarrow{A%
}_{g}=\frac{G}{c}\left[ \frac{3\widehat{r}(\widehat{r}\cdot \overrightarrow{J%
})-\overrightarrow{J}}{r^{3}}\right]   \label{bj}
\end{equation}
It is interesting to see that the condition $\overline{h}^{\mu \nu },_{\mu
}=0$ leads to ${\bf \nabla }\cdot \overrightarrow{A}_{g}=0$ (analogous to
the Coulomb gauge of electromagnetism).

Let us note that for the case of a slowly rotating sphere with angular
momentum $\overrightarrow{J}=(0,0,J)$, we obtain from (\ref{ag}) in
spherical coordinates 
\begin{equation}
\overline{h}_{0\varphi }=h_{0\varphi }=-\frac{2GJ}{c^{3}r}\sin ^{2}\theta
\label{kerr}
\end{equation}
which is the $g_{0\varphi }$ component of the Kerr metric in Boyer-Lindquist
coordinates in the weak field and slow motion limit\cite{Boyer}.

\section{The weak field approximation of Brans-Dicke theory}

\bigskip In Brans-Dicke theory of gravity the field equations are given by 
\cite{Brans} 
\begin{equation}
G_{\mu \upsilon }=\frac{8\pi }{c^{4}\phi }T_{\mu \upsilon }+\frac{\omega }{%
\phi ^{2}}(\phi ,_{\mu }\phi ,_{\upsilon }-\frac{1}{2}g_{\mu \upsilon }\phi
,_{\alpha }\phi ^{,\alpha })+\frac{1}{\phi }(\phi ,_{\mu ;\upsilon }-g_{\mu
\upsilon }\square \phi )  \label{exatas}
\end{equation}

Paralleling general relativity one can linearize Brans-Dicke field equations
by assuming that the metric $g_{\mu \nu }$ and the scalar field $\phi $ can
be written as $g_{\mu \nu }=\eta _{\mu \nu }+h_{\mu \upsilon }$ and $\phi
=\phi _{0}+\epsilon $, where $\phi _{0}$ is a constant and $\epsilon
=\epsilon (x)$ is a first-order term ( it is assumed that both $\left|
h_{\mu \upsilon }\right| $ and $\left| \epsilon \phi _{0}^{-1}\right| $ are $%
\ll 1$). From these assumptions it follows that 
\begin{equation}
\square h_{\mu \nu }=-\frac{16\pi }{c^{4}\phi _{0}}\left[ T_{\mu \upsilon }-%
\frac{\omega +1}{2\omega +3}\eta _{\mu \nu }T\right]   \label{BDlinear}
\end{equation}
where we have used the Brans-Dicke gauge\cite{Brans} $\left( h_{\upsilon
}^{\mu }-\frac{1}{2}\delta _{\nu }^{\mu }h\right) ,_{\mu }=\epsilon
,_{\upsilon }\phi _{0}^{-1}$.

It has been shown that the problem of finding solutions of Brans-Dicke
equations of gravity in the weak field approximation \ may be reduced to
solving the linearized Einstein field equations for the same energy-momentum
tensor\cite{Barros}. Indeed, if $g_{\mu \nu }^{\ast }(G,x)$ is a known
solution of the Einstein equations in the weak field approximation for a
given $T_{\mu \upsilon }$, then the Brans-Dicke solution corresponding to
the same $T_{\mu \upsilon }$ will be given in the weak field approximation
by 
\begin{equation}
g_{\mu \upsilon }(x)=[1-\epsilon G_{0}]g_{\mu \nu }^{\ast }(G_{0},x)
\label{go}
\end{equation}
where $G$ is the gravitational constant and $G_{0}=\phi _{0}^{-1}=\left( 
\frac{2\omega +3}{2\omega +4}\right) G$ and the function $\epsilon (x)$ is a
solution of the scalar field equation 
\begin{mathletters}
\begin{equation}
\square \epsilon =\frac{8\pi T}{c^{4}(2\omega +3)}  \label{traço}
\end{equation}
$T$ denoting the trace of $T_{\mu \upsilon }$.

\section{The gravitomagnetic field in Brans-Dicke theory}

Let us now proceed to the definition of the gravitomagnetic field in
Brans-Dicke theory. We start by defining $\overline{h}_{\mu \nu }$ as 
\end{mathletters}
\begin{equation}
\overline{h}_{\mu \nu }=h_{\mu \upsilon }-\frac{1}{2}\eta _{\mu \upsilon
}h-\epsilon G_{0}\eta _{\mu \upsilon }  \label{gaugebdi}
\end{equation}
It is easily seen from (\ref{BDlinear}) that 
\begin{equation}
\square \overline{h}_{\mu \nu }=-\frac{16\pi G_{0}}{c^{4}}T_{\mu \nu }
\label{gobd}
\end{equation}
Thus, in close analogy to the general relativity case, if we again restrict
ourselves to slow motion and stationary sources, we immediately arrive at
the following equations, which are supposed to hold far from the source 
\[
\overline{h}_{00}\equiv \frac{4\Phi _{g}^{BD}}{c^{2}}=\frac{4G_{0}M}{c^{2}r} 
\]
\[
\overrightarrow{\overline{h}}=-\frac{2G_{0}(\overrightarrow{J}\times 
\overrightarrow{r})}{c^{3}r^{3}}\equiv -\frac{2\overrightarrow{A}_{g}^{BD}}{%
c^{2}} 
\]

Therefore, in the context of Brans-Dicke theory we define the
gravitoelectric field to be $\overrightarrow{E_{g}}^{BD}=-{\bf \nabla }\Phi
_{g}^{BD}$ and the gravitomagnetic field to be 
\begin{equation}
\overrightarrow{B}_{g}^{BD}=\overrightarrow{{\bf \nabla }}\times 
\overrightarrow{A}_{g}^{BD}=\frac{G_{0}}{c}\left[ \frac{3\widehat{r}(%
\widehat{r}\cdot \overrightarrow{J})-\overrightarrow{J}}{r^{3}}\right]
\label{bgb}
\end{equation}

Another way of deriving the results above is to start with the Kerr metric
in the weak field and slow motion approximation given in isotropic
coordinates 
\[
ds^{2}=-\left( 1-\frac{2GM}{c^{2}r}\right) c^{2}dt^{2}+\left( 1+\frac{2GM}{%
c^{2}r}\right) \left( dr^{2}+r^{2}(d\theta ^{2}+\sin ^{2}\theta d\varphi
^{2})\right) -\frac{4GJ}{c^{3}r}\sin ^{2}\theta d\varphi cdt
\]
Following the prescription given by the theorem discussed in Section III the
corresponding solution in Brans-Dicke theory in the same approximation will
be given by 
\begin{eqnarray*}
ds^{2} &=&[1-\epsilon G_{0}][-\left( 1-\frac{2G_{0}M}{c^{2}r}\right)
c^{2}dt^{2}+\left( 1+\frac{2G_{0}M}{c^{2}r}\right) \left(
dr^{2}+r^{2}(d\theta ^{2}+\sin ^{2}\theta d\varphi ^{2})\right)  \\
&&-\frac{4G_{0}J}{c^{3}r}\sin ^{2}\theta d\varphi cdt]
\end{eqnarray*}
with the function $\epsilon $ satisfying the equation (\ref{traço}) (recall
that in this case there is no contribution from the angular momentum of the
rotating body to the trace of the energy-momentum tensor). Therefore, for a
slowly rotating sphere with angular momentum $\overrightarrow{J}=(0,0,J)$ we
will have 
\begin{equation}
\overline{h}_{0\varphi }=h_{0\varphi }=-\frac{2G_{0}J}{c^{3}r}\sin
^{2}\theta   \label{hoif}
\end{equation}

\section{The Lense-Thirring effect in Brans-Dicke theory}

As is well known, the Lense-Thirring effect consists in a precession of
gyroscopes relative to distant stars, or, equivalently, a dragging of
inertial frames, an effect caused by the gravitomagnetic field. Denoting the
angular momentum and the angular velocity of the precession by $%
\overrightarrow{S}$ and $\overrightarrow{\Omega }$, then the torque acting
on the gyroscope predicted by general relativity is given by 
\begin{equation}
\overrightarrow{\tau }=\frac{1}{2}\overrightarrow{S}\times \left( -\frac{2}{%
c^{2}}\overrightarrow{B}_{g}\right) =\frac{d\overrightarrow{S}}{dt}=%
\overrightarrow{\Omega }\times \overrightarrow{S}  \label{precession}
\end{equation}
with 
\begin{equation}
\overrightarrow{\Omega }=\frac{1}{c^{2}}\overrightarrow{B}_{g}=G\left( \frac{%
3\widehat{r}(\widehat{r}\cdot \overrightarrow{J})-\overrightarrow{J}}{%
c^{3}r^{3}}\right)   \label{omega}
\end{equation}
Thus in the case of Brans-Dicke theory the equation (\ref{omega})\ becomes 
\begin{equation}
\overrightarrow{\Omega }^{BD}=\frac{1}{c^{2}}\overrightarrow{B}%
_{g}^{BD}=G_{0}\left( \frac{3\widehat{r}(\widehat{r}\cdot \overrightarrow{J}%
)-\overrightarrow{J}}{c^{3}r^{3}}\right)   \label{omegao}
\end{equation}

To compare the value of $\Omega $ predicted by general relativity with $%
\Omega ^{BD}$, predicted by Brans-Dicke theory, we must ascribe values for $%
\omega $, the scalar field coupling constant. According to the latest
experimental results (VLBI measurements \cite{Will} ) the current value for $%
\omega $ is $3500$ . On the other hand, for a polar orbit at about $650$ km
altitude the axis of a gyroscope \ is predicted to undergo a precession rate
of $42$ milliarcsec per year . The expected accuracy of the experiment under
these conditions (Gravity Probe B) is about $0.5$ milliarcsec per year.
Since $G_{0}=\left( \frac{2\omega +3}{2\omega +4}\right) G$ the predicted
value of Brans-Dicke theory is 
\[
\Omega ^{BD}=\frac{7003}{7004}\Omega \simeq 41.99\text{ milliarcsec per year.%
}
\]
Therefore we see that within the precision of the experiment one cannot
distinguish one theory from another.

\section{The gravitomagnetic time delay in Brans-Dicke theory}

The time delay of light is considered a classical test of general relativity
and its measurement was first proposed by Shapiro \cite{Shapiro}. It can be
shown that this effect can be separated into two parts: the Shapiro time
delay and the gravitomagnetic time delay, the latter due to the
gravitomagnetic field. The gravitomagnetic time delay has been investigated
recently by Ciufolini et al \cite{ciufolini}. Assuming again the weak field
and slow motion approximation of general relativity one can show that the
gravitational time delay $\Delta $ of a light signal travelling between two
points $P_{1}$ and $P_{2}$ is given by 
\begin{equation}
\Delta =\frac{1}{2c}\int_{P_{1}}^{P_{2}}\overline{h}_{\mu \upsilon
}(x)k^{\mu }k^{\nu }dl  \label{delta}
\end{equation}
where $k^{\mu }=(1,\widehat{k})$, $\widehat{k}$ denotes the light
propagation unit vector and $dl=\left| d\overrightarrow{r}\right| $ is the
Euclidean length element along the straight line that joins $P_{1}$ to $%
P_{2}.$ Now from (\ref{ro}), (\ref{ag}) and (\ref{delta}) it follows that $%
\Delta =\Delta _{ge}+\Delta _{gm}$, where 
\begin{equation}
\Delta _{ge}=\frac{2}{c^{3}}\int_{P_{1}}^{P_{2}}\Phi _{g}dl
\label{shapirotime}
\end{equation}
\ is the Shapiro delay and 
\begin{equation}
\Delta _{gm}=-\frac{2}{c^{3}}\int_{P_{1}}^{P_{2}}\overrightarrow{A}_{g}\cdot
d\overrightarrow{r}  \label{gtd}
\end{equation}
is the gravitomagnetic time delay.

Clearly, the above equations keep exactly the same form when we go from
general relativity to Brans-Dicke theory, the only change needed is the
substitution $\Phi _{g}$ $\rightarrow $ $\Phi _{g}^{BD}$ and $%
\overrightarrow{A}_{g}$ $\rightarrow \overrightarrow{A}_{g}^{BD}$. Thus we
have 
\begin{equation}
\Delta _{ge}^{BD}=\left( \frac{2\omega +3}{2\omega +4}\right) \Delta _{ge}
\label{rel1}
\end{equation}
\begin{equation}
\Delta _{gm}^{BD}=\left( \frac{2\omega +3}{2\omega +4}\right) \Delta _{gm}
\label{rel2}
\end{equation}

At this point two comments are in order. Firstly, it should be noted that
analogously to the general relativity approach the gravitomagnetic echo
delay vanishes \cite{ciufolini}. Secondly, if the light rays travel along a
closed loop around a rotating body (this can be arranged with the help of
``mirrors''), then the time delay due to the gravitomagnetic field depends
on the direction the rays go around the loop. Similarly to the general
relativity case, the total time diference between two opposite-oriented
paths is given by 
\[
\delta t^{BD}=-\frac{4}{c^{3}}\oint \overrightarrow{A}_{g}^{BD}\cdot d%
\overrightarrow{r}=-\frac{4}{c^{3}}\left( \frac{2\omega +3}{2\omega +4}%
\right) \oint \overrightarrow{A}_{g}\cdot d\overrightarrow{r} 
\]

\section{Final remarks}

In this article, we have examined two effects associated with the so-called
gravitomagnetism, namely, the Lense-Thirring effect and the gravitomagnetic
time delay in Brans-Dicke theory of gravity. Following the same line of
reasoning employed in this article it can easily be shown that the equations
for the gravitational time delay  in different images due to gravitational
lensing\cite{ciufolini} in Brans-Dicke theory may be obtained again from the
corresponding equations in general relativity by using the correction factor 
$\frac{2\omega +3}{2\omega +4}.$ We believe that subsequent analyses of
other weak field effects where the gravitomagnetic field can play a role may
benefit of the simple theorem of Section III, which establishes a direct
correspondence between general relativity and Brans-Dicke theory of gravity
in the weak field approximation.

It is interesting to note that a major motivation that has led to the
formulation of Brans-Dicke theory was the quest for a Machian theory of
gravity\cite{Brans}. In this respect, one could say that the prediction of
the gravitomagnetism, or the dragging of inertial frames, by both general
relativity and Brans-Dicke theory alike make these theories, at least in a
``weak sense'' (as pointed out by Ciufolini et al \cite{lageos}), satisfy
the Principle of Mach.

\qquad

\end{document}